\documentstyle[twoside,fleqn,espcrc2]{article}
\title{Uniform Twistor--Like Formulation of
Massive and Massless Superparticles with Tensorial Central
Charges}

\author{S. Fedoruk\address{Ukrainian
Engineering--Pedagogical Academy, \\
61003 Kharkiv, 16 Universitetska Str., Ukraine \\
e-mail: fed@postmaster.co.uk} and
V. G. Zima\address{Kharkiv National University, \\
61077 Kharkiv, 4 Svobody Sq., Ukraine \\
e-mail: zima@postmaster.co.uk}}

\begin{document}

\begin{abstract}
We construct the manifestly Lorentz-invariant twistorial
formulation of the $N=1$ $D=4$ superparticle with tensorial
central charges which describes massive and massless cases in a
uniform manner. The tensorial central charges are realized in
terms of even spinor variables and central charge coordinates. The
full analysis of the number of conserved supersymmetries has been
carried out. In the massive case the superparticle preserves
$1/4$ or $1/2$ of target-space supersymmetries whereas the massless
superparticle preserves two or three supersymmetries.
\vspace{1pc}
\end{abstract}

\maketitle

\section{Introduction}
In a recent paper~\cite{FZ} we proposed
a new relativistic formulation of massive superparticle with
tensorial central charges~\cite{AzG}-\cite{GaHu}. The model
contains a commuting Weyl spinor as a collection of coordinates
of the configuration space and describes a superparticle whose
presence breaks two or three of
 $N=1$, $D=4$ target-space supersymmetries. It is interesting that
in the background of central charges the massive superparticle is
equivalent to massive spinning particle ~\cite{Pseudo},
\cite{Pseu} if a quarter of target-space supersymmetry is
preserved. In a certain sense the commuting spinor variables of
the model play the role of index spinor
variables~\cite{ZimFedL}-\cite{ZimFedJ}. This model does not
contain any special coordinates for the tensorial central charges.
Analogous model of massive superparticle preserving
$1/4$ of target-space supersymmetries has been formulated
in~\cite{DelIvKr} without explicit Lorentz covariance.

It should be mentioned that D. V. Volkov and his collaborators
have proposed one of the first twistor-like models for the
massless superparticle~\cite{STV} and established the equivalence
between the spinning particle and the usual superparticle without
central charges at least on the classical level. The idea of
identifying the $\kappa$-symmetry of the superparticle with the
local worldline supersymmetry of the spinning particle has been a
basic one for the superfield formulation of massless superparticle
theory~\cite{STV} and its generalization to the superembedding
description of superbranes~\cite{BSV}.

In this paper we present a twistorial formulation of the
superparticle with tensorial central charges in which massive and
massless cases are described in uniform manner. The model uses
both the central charge coordinates and  the auxiliary bosonic
spinor variables simultaneously. Due to the use of spinors the
analysis is simplified by reducing the tensorial quantities to
scalar ones. For zero mass our model reduces to the twistorial
formulation of the massless superparticle with tensorial central
charges~\cite{BandL} in which one or two of target-space
supersymmetries are broken. In the massive case we have a
bitwistorial formulation of the massive superparticle with
tensorial central charges preserving $1/4$ or $1/2$ of
target-space supersymmetries.

\section{The formulation of the model}

The configuration space of the model is parametrized along with
the usual superspace coordinates $x^\mu$, $\theta^\alpha$,
$\bar\theta^{\dot\alpha}$ and the tensorial central charge
coordinates
$y^{\alpha\beta} =y^{(\alpha\beta )}$,
$\bar y^{\dot\alpha\dot\beta} =\bar y^{(\dot\alpha\dot\beta )} =
\overline{(y^{\alpha\beta})}$ also by two bosonic spinors
$v_\alpha{}^a$, $\bar v_{\dot\alpha{}a} =\overline{(v_\alpha{}^a)}$,
$a=1,2$, $\alpha =1,2$, $\dot\alpha =1,2$.
Weyl spinor indices are raised or lowered with the help of the
unit invariant skew-symmetric matrices $\epsilon^{\alpha\beta}
=\epsilon^{\dot\alpha\dot\beta}$ and $\epsilon_{\alpha\beta}
=\epsilon_{\dot\alpha\dot\beta}$, i. e. $A^\alpha
=\epsilon^{\alpha\beta} A_\beta$,
$A_\alpha =\epsilon_{\alpha\beta} A^\beta$,
$\bar B^{\dot\alpha} =\epsilon^{\dot\alpha\dot\beta}\bar B_{\dot\beta}$,
$\bar B_{\dot\alpha} =\epsilon_{\dot\alpha\dot\beta}\bar B^{\dot\beta}$
for arbitrary $A$ and $\bar B$. We use $D=4$  Weyl spinor  and
$\sigma$--matrices  conventions of~\cite{Wess}. The metric tensor
$\eta_{\mu\nu}$ signature is mostly plus and
$\sigma_{\alpha\dot\beta}^{(\mu}\tilde\sigma^{\nu
)\dot\beta\gamma} =-\delta^\gamma_\alpha \eta^{\mu\nu}  $. It will
be convenient to write the relations of the model formally as
$SU(2)$--covariants with respect to the indices $a$, $b$, $c$,
... numbering bosonic spinors. So these indices are raised or
lowered as
$SU(2)$ ones by the matrices $\epsilon^{ab} = -\epsilon_{ab}$
and their positions are interchanged under the complex conjugation.

For description of the superparticle
with tensorial central charges we take the action in
twistor-like form
\begin{eqnarray} \label{act}
S&=&\int d\tau \, L \, ,\nonumber \\
 L & = & P_\mu \Pi^\mu_\tau
+Z_{\alpha\beta}\Pi^{\alpha\beta}_\tau +\bar
Z_{\dot\alpha\dot\beta}\bar\Pi^{\dot\alpha\dot\beta}_\tau
- \nonumber \\
& & \lambda (v^{\alpha{}a}v_{\alpha{}a} -2m) -
\bar\lambda (\bar v_{\dot\alpha{}a}\bar v^{\dot\alpha{}a} -2m) \, .
\label{Lagr}
\end{eqnarray}
Here  the one-forms
\begin{eqnarray}\nonumber
\Pi^{\mu}&\equiv& d\tau \Pi^{\mu}_\tau =
dx^\mu -id\theta\sigma^\mu\bar\theta + i\theta\sigma^\mu d\bar\theta \, ,\\
 \label{form}
\Pi^{\alpha\beta}&\equiv& d\tau \Pi^{\alpha\beta}_\tau =
dy^{\alpha\beta} + i\theta^{(\alpha} d\theta^{\beta )} \, ,\\
\nonumber
\bar\Pi^{\dot\alpha\dot\beta}&\equiv & d\tau \bar\Pi^{\dot\alpha\dot\beta}_\tau
= d\bar y^{\dot\alpha\dot\beta} +
i\bar\theta^{(\dot\alpha} d\bar\theta^{\dot\beta )}
\end{eqnarray}
are invariant  under global  supersymmetry transformations
\begin{eqnarray}\nonumber
\delta\theta^\alpha& =&\epsilon^\alpha \, , \quad
\delta\bar\theta^{\dot\alpha} =\bar\epsilon^{\dot\alpha} \, ,\\
\nonumber
\delta x^\mu& =& i\theta\sigma^\mu \delta \bar\theta -
i\delta \theta\sigma^\mu \bar\theta \,  ,\\
\nonumber
\delta y^{\alpha\beta}& =& i\theta^{(\alpha}\delta\theta^{\beta )} \, , \quad
\delta\bar y^{\dot\alpha\dot\beta} =
i\bar\theta^{(\dot\alpha} \delta\bar\theta^{\dot\beta )}  \,  ,\\
  \label{glsus}
\delta v_{\alpha}{}^a& =&0  \, , \quad
\delta \bar v_{\dot\alpha{}a}=0
\end{eqnarray}
acting in the extended  superspace parametrized by the usual
superspace coordinates $x^\mu$, $\theta^\alpha$,
$\bar\theta^{\dot\alpha}$ and by the tensorial central charge
coordinates $y^{\alpha\beta}$,
$\bar y^{\dot\alpha\dot\beta}$.

The  quantities
$P_\mu$, $Z_{\alpha\beta}=Z_{\beta\alpha}$,
$\bar Z_{\dot\alpha\dot\beta}=\bar Z_{\dot\beta\dot\alpha}$,
which play  the  role of  the momenta for $x^\mu$, $y^{\alpha\beta}$,
$\bar y^{\dot\alpha\dot\beta}$, are taken as the sums of products
of two bosonic spinors $v_{\alpha}{}^a$, $\bar v_{\dot\alpha{}a}$
\begin{equation} \label{P}
P_{\alpha\dot\beta} =P_\mu \sigma^\mu_{\alpha\dot\beta} =
v_{\alpha}{}^a \bar v_{\dot\alpha{}a}\, ,
\end{equation}
\begin{equation} \label{Z}
Z_{\alpha\beta} = v_{\alpha}{}^a v_{\beta}{}^b C_{ab} \, ,
\end{equation}
\begin{equation} \label{barZ}
\bar Z_{\dot\alpha\dot\beta} =
\bar v_{\dot\alpha{}a}\bar v_{\dot\beta{}b} \bar C^{ab} \, ,
\end{equation}
where $C_{ab}$, $\bar C^{ab} =\overline{(C_{ab})}$ are symmetric
constant matrices. These expressions are completely general with
respect to the four--momentum $P_{\alpha\dot\beta}$ but imply
some constraints on the central charges
$Z_{\alpha\beta}$, $\bar Z_{\dot\alpha\dot\beta}$. Here we do not give
the explicit formulation of these constraints.

Due to the kinematic constraints
\begin{equation}
v^{\alpha{}a}v_{\alpha{}a} =2m  \, ,  \qquad
\bar v_{\dot\alpha{}a}\bar v^{\dot\alpha{}a} =2m  \, ,
\end{equation}
which are  equivalent to
\begin{equation} \label{kinem}
v^{\alpha{}a}v_{\alpha{}b} =m\delta^a_b  \, ,  \qquad
\bar v_{\dot\alpha{}a}\bar v^{\dot\alpha{}b} =m\delta^b_a  \, ,
\end{equation}
and enter the action~(\ref{act}) with Lagrange multipliers
we have ${\rm det}(v_{\alpha}{}^a)=m$ and
\begin{equation}
P^2  \equiv P^\mu P_\mu =-m^2  \, .
\end{equation}
Thus the constant $|m|$ plays the role of the  mass. It should be
noted that the change of  the sign of $m$ is equivalent to
antipodal transformations $v_\alpha ^1 \leftrightarrow v_\alpha
^2$ of bosonic spinors in ``internal space'' which leaves
invariant the quadratic expressions~(\ref{P})-(\ref{barZ}) for
the energy--momentum vector and  central charges of the model.

In the massless  case ($m=0$) the spinors $v_{\alpha}{}^1$ and
$v_{\alpha}{}^2$ are proportional to each other $v_{\alpha}{}^1 \sim
v_{\alpha}{}^2$ as the consequense of the kinematic
constraints~(\ref{kinem}). As a result one obtains a formulation
of massless superparticle with one bosonic spinor from which both
the massless four--momentum and the tensorial central charge are
constructed. Such a model has been analyzed in~\cite{BandL}. The
number of preserved SUSY is equal two or three in this model. In
the proposed model~(\ref{act}) we use a minimal number of bosonic
spinors, which is two, for constructing the energy-momentum vector
with arbitrary mass~\cite{Penr}. Therefore we regard our
formulation as the twistor--like one and concentrate on the
massive case in the following.

Coefficients in the expansion of the symmetric central charge
matrix $C_{ab}$ in terms of the Pauli matrices
$(\sigma_i)_{a}{}^{b}$ form a
 complex dimensionless
``internal'' three--vector ${\bf C} =i({\bf E} +i{\bf H})$,
real and imaginary parts
of which we denote by analogy with electrodynamics. Thus
\begin{equation}
C_{ab}={\rm C}_{i}(\sigma_i)_{ab} \, ,\quad
\bar C^{ab}=-\bar{\rm C}_{i}(\sigma_i)^{ab}
\end{equation}
and
\begin{equation}\nonumber
C_{ab}C^{bc}=-{\bf C}{\bf C}\delta_a^c = ({\bf E}^2-{\bf H}^2
+2i{\bf E}{\bf H})\delta_a^c,
\end{equation}
\begin{equation}\nonumber
C_{ab}\bar C^{ab}=2{\bf C}\bar{\bf C}=2({\bf E}^2+{\bf H}^2) \, .
\end{equation}

One can simplify the matrix $C$ of the  central charges using
redefinitions of the bosonic spinors with unitary unimodular
transformation acting on the  indices $a$, $b$, ... and leaving
intact the four--momentum matrix and kinematic constraints. In
fact with some loss of generality we could take the matrix $C$ to
be diagonal from the beginning.

\section{$\kappa$--symmetry transformations}

The  variations of bosonic
coordinates under the  local $\kappa$--symmetry
 transformations~\cite{AsLuk}, \cite{Sieg}
  has the same form in terms of the variations
of odd spinor coordinates as SUSY variations but are opposite in
sign
\begin{eqnarray}
\delta x^\mu& =& -i\theta\sigma^\mu \delta \bar\theta +
i\delta \theta\sigma^\mu \bar\theta \,  ,\nonumber\\
\delta y^{\alpha\beta}& =& -i\theta^{(\alpha}\delta\theta^{\beta )} \, , \quad
\delta\bar y^{\dot\alpha\dot\beta} =
-i\bar\theta^{(\dot\alpha} \delta\bar\theta^{\dot\beta )}  \,  ,\\
\delta v_\alpha{}^a& =&0  \, , \quad
\delta \bar v_{\dot\alpha{}a} =0  \, .\nonumber
\end{eqnarray}
Further, for the one--forms~(\ref{form}) in the action we have
\begin{equation}
\delta \Pi^\mu_\tau = -2i\dot\theta\sigma^\mu \delta \bar\theta +
2i\delta \theta\sigma^\mu \dot{\bar\theta} \,  ,
\end{equation}
\begin{equation}
\delta \Pi_\tau^{\alpha\beta} =
-2i\dot\theta^{(\alpha}\delta\theta^{\beta )} \, , \quad
\delta\bar \Pi_\tau^{\dot\alpha\dot\beta} =
-2i\dot{\bar\theta}^{(\dot\alpha} \delta\bar\theta^{\dot\beta )}\, .
\end{equation}
The
corresponding variation of the Lagrangian is
\begin{eqnarray}
\delta L & = &
2i(\bar v_a\delta\bar\theta +
C_{ab}\delta\theta v^b)\dot\theta v^a \nonumber \\
& & + 2i(\delta\theta v^a +\bar C^{ab}\bar v_b\delta\bar\theta)
\bar v_a\dot{\bar\theta}  \, .
\end{eqnarray}

The most general variations of the Grassmann spinors under
$\kappa$--symmetry are
\begin{equation} \label{loctr}
\delta\theta^\alpha =\kappa_a v^{\alpha{}a} \, ,  \quad
\delta\bar\theta^{\dot\alpha} =\bar\kappa^a \bar v^{\dot\alpha}{}_a
\end{equation}
with two complex local Grassmann parameters $\kappa_a(\tau)$,
$\bar\kappa^a(\tau) =\overline{(\kappa_a )}$. Taking into account  the
normalization conditions for the bosonic spinors~(\ref{kinem}) we
arrive at
\begin{eqnarray}
\delta L & = &
2im(\bar\kappa^a + C^{ab}\kappa_b)\dot\theta v^a
- \nonumber \\
& & 2im(\kappa_a +\bar C_{ab}\bar\kappa^b)\bar v^a\dot{\bar\theta}  \, .
\end{eqnarray}

The number of preserved supersymmetries is defined by the number
of independent functions $\kappa_a$, $\bar\kappa^a$ for which
$\delta L=0$. Hence the equations
\begin{equation} \label{k}
\kappa_a +\bar C_{ab}\bar\kappa^b =0 \, ,\quad
\bar\kappa^a + C^{ab}\kappa_b=0
\end{equation}
should have nontrivial solutions when there is $\kappa$--symmetry.
These equations can be written in the matrix form
\begin{equation} \label{matk}
\Delta \,{\cal K} =0
\end{equation}
where
\begin{equation}
\Delta =\left(
\begin{array}{cc}
\delta_a^{b} & \bar C_{ab}  \\
C^{ab} & \delta^{a}_b
\end{array}
\right)  \,
\quad \mbox{and} \quad
{\cal K} = \left( \kappa_a \atop \bar\kappa^a \right) \, .
\end{equation}
The matrix  $\Delta$ is Hermitian,
$\Delta =\Delta^+$, therefore it is unitary diagonalizable.
The number of the independent $\kappa$--symmetries (solutions of
eqs.~(\ref{k})) coincides with the number of the zero eigenvalues
of the matrix $\Delta$.

One can easily obtain that
\begin{equation}
{\rm det}\Delta = 1- C^{ab}\bar C_{ab} +
\frac{1}{4}C^{ab}C_{ab}\bar C^{cd}\bar C_{cd} \, .
\end{equation}
So the necessary condition for the presence of
$\kappa$--symmetries (one or more) consists in equality
\begin{equation} \label{pres}
{\rm det}\Delta =0\, .
\end{equation}

Some algebra gives
\begin{eqnarray}
&&{\rm det}(\Delta -\lambda {\bf 1}_4)  =  \nonumber \\
& = & \Lambda^4 -
\Lambda^2 C^{ab}\bar C_{ab} +
\frac{1}{4}C^{ab}C_{ab}\bar C^{cd}\bar C_{cd}  \nonumber \\
& = & (\Lambda^2 -{\bf E}^2 -{\bf H}^2)^2
- 4|{\bf E} \times{\bf H}|^2
\end{eqnarray}
with $\Lambda \equiv 1-\lambda$.

The characteristic equation reads
\begin{equation} \label{lambda}
\lambda^4 -4\lambda^3 -k_2 \lambda^2 +2k_1 \lambda +k_0 =0
\end{equation}
where the coefficients are
\begin{eqnarray}
k_2 =C^{ab}\bar C_{ab}-6=2({\bf E}^2 +{\bf H}^2 -3) \, ,\nonumber\\
k_1 =C^{ab}\bar C_{ab}-2=2({\bf E}^2 +{\bf H}^2 -1) \, ,\nonumber\\
k_0 ={\rm det}\Delta  = ({\bf E}^2 +{\bf H}^2 -1)^2  - 4|{\bf E}
\times{\bf H}|^2\, .\nonumber
\end{eqnarray}
Let us now consider all possible eigenvalues of the matrix
$\Delta$.

\subsection{$3/4$ unbroken SUSY}
The presence  of three zero  eigenvalues means  that the
characteristic equation~(\ref{lambda}) must be of the form
$$
\lambda^3 (\lambda -\lambda_1 )=0 \, .
$$
This gives us the conditions $k_2 =k_1 =k_0 =0$ on the
coefficients of eq.~(\ref{lambda}). However as one can see from
the explicit expressions for
 the coefficients in our model the inequality $k_1 \ne k_2$ is always fulfilled.
 Therefore the presence of  three zero eigenvalues is  not
possible in the massive case of the model under consideration. So
one can not get three first class fermionic constraints and
$3/4$ unbroken SUSY in this case.

\subsection{$1/2$ unbroken SUSY}
For two zero eigenvalues or $1/2$ unbroken SUSY the equation on
the eigenvalues
$\lambda^2 (\lambda -\lambda_1 )(\lambda -\lambda_2 )=0$ means that
$k_1 =k_0 =0$ in eq.~(\ref{lambda}).
This gives us two conditions  on parameters of the central charges
$$
C^{ab}\bar C_{ab}=2 \, , \quad C^{ab}C_{ab}\bar C^{cd}\bar
C_{cd}=4
\,
$$
or equivalently in the $3$--vector form
$$
{\bf E}^2 +{\bf H}^2 =1 \, ,
$$
$$
{\bf E} \times{\bf H} = 0 \, .
$$
Thus in this case the vectors ${\bf E}$ and ${\bf H}$ are
parallel, and they are not equal to zero simultaneously. If two
eigenvalues are zero then two nonzero eigenvalues are both equal
to $2$.

Note that the above conditions, which define the case with $1/2$
ubrokne SUSY, are equivalent to
$$
C^{ab}\bar C_{ab}=2 \, , \quad C^{ab}\bar C_{bc}C^{cd}\bar
C_{da}=2
\,
$$
which are obtained by the Fierz transformation
$$
C^{ab}C_{ab}\bar C^{cd}\bar C_{cd}=2(C^{ab}\bar C_{ab})^2
-2C^{ab}\bar C_{bc}C^{cd}\bar C_{da}\, .
$$
Due to the first condition
$C^{ac}\bar C_{cb}=\delta^a{}_b +A^a{}_b$, the
 matrix  $A$ is traceless,
$A^b{}_b=0$ and Hermitian, $A^+=A$. But due to the second condition
we obtain the equation $A^a{}_bA^b{}_a=0$ which gives us
$A^a{}_b=0$. Thus in the case of two $\kappa$--symmetries ($1/2$
 SUSY preserved) the coefficient matrix of the central charges is unitary
\begin{equation}\label{ortC}
C^{ac}\bar C_{cb}=\delta^a{}_b \, .
\end{equation}

The solutions of eqs.~(\ref{k}), provided that
condition~(\ref{ortC}) is fulfilled, can be obtained after the
diagonalization of the matrix $\Delta$
\begin{equation} \label{delt}
\Delta_{\rm diag}\equiv \left(
\begin{array}{cc}
0\cdot {\bf 1}_2 & 0  \\
0 & 2\cdot {\bf 1}_2
\end{array}
\right) =V\Delta V^{-1}\, ,
\end{equation}
with
$$
V =\frac{1}{\sqrt{2}} \left(
\begin{array}{cc}
\delta_a{}^b & -\bar C_{ab}  \\
C^{ab} & \delta^a{}_b
\end{array}
\right) \, .
$$
To verify the unitarity of the matrix $V$ and the
equality~(\ref{delt}) we have used the condition~(\ref{ortC}).
Thus eq.~(\ref{k}) takes a simple form
\begin{equation} \label{kap}
\Delta_{\rm diag} \,{\cal K}^\prime =0
\end{equation}
where ${\cal K}^\prime =V{\cal K}$. Obviously the solution
of eq.~(\ref{kap}) is
$$
{\cal K}^\prime = \left( \nu_a \atop 0 \right) \, .
$$
However the condition of mutual conjugacy
$\bar\kappa^a =\overline{\kappa_a}$ of the upper and lower part of
the column ${\cal K}$ should be taken into account. To this end
let us represent the symmetric unitary matrix $C$ as a square of a
symmetric unitary matrix $\sqrt{C}$, whose explicit form is not
required. Then for an arbitrary real odd two-component quantity
$\rho$ the quantity $\nu = \sqrt{C}
\rho$ satisfies the required conjugation condition. Thus we have
demonstrated that the parameter space of the
$\kappa$--transformations is actually a two--dimensional real space.

Eigenvectors corresponding to the eigenvalue $2$ can be obtained
in the similar way. But now
${\cal K}^\prime = \left( 0 \atop \bar\nu^\prime \right)$ where
$\bar\nu^\prime = \sqrt{C} \rho^\prime$ and this space is parameterized
by two arbitrary real odd quantities collected in the
two-component "vector"
$\rho^\prime$.

\subsection{$1/4$ unbroken SUSY}
For a single zero eigenvalue or for $1/4$ unbroken SUSY we have
the single condition
$$
k_0 ={\rm det}\Delta = 1- C^{ab}\bar C_{ab} +
\frac{1}{4}C^{ab}C_{ab}\bar C^{cd}\bar C_{cd} =0\,
$$
which in term of the vectors ${\bf E}$ and ${\bf H}$ has the form
$$
|{\bf E}^2 +{\bf H}^2 -1| =
2|{\bf E} \times{\bf H}| \, .
$$
In this  case the  characteristic  equation is
\begin{eqnarray}
&&\lambda (\lambda -2)(\lambda -1 -\sqrt{C^{ab}\bar C_{ab}-1}) \times
  \nonumber \\
&& (\lambda -1 +\sqrt{C^{ab}\bar C_{ab}-1})  =0  \nonumber
\end{eqnarray}
and the three nonzero eigenvalues  are $\lambda =2$,
$\lambda =1\pm\sqrt{C^{ab}\bar C_{ab}-1}$.

As it has been noted above the arbitrary symmetric matrix
$C$ can be reduced
to the diagonal form
$$
C^{\prime} = \left(
\begin{array}{cc}
\rho_1 e^{i\varphi_1} & 0  \\
0 & \rho_2 e^{i\varphi_2}
\end{array}
\right) = VCV^{-1}
$$
by means of the ``internal'' $SU(2)$--transformation $V$. Here
$\rho_1$, $\rho_2$ and $\varphi_1$, $\varphi_2$ are real.
One can easily obtain that
$\rho_{1,2}^2 = {\bf E}^2 +{\bf H}^2 \pm |{\bf E}^2 \times{\bf H}^2|$.
The case when $\rho_1 =\rho_2 =1$ and the matrix $C$ is unitary
has been considered in the  previous subsection. Now we have
$$
C^{ab}\bar C_{ab}=\rho_1^2 +\rho_2^2
$$
and
$$
{\rm det} \Delta =(\rho_1^2 -1)(\rho_2^2 -1) \, .
$$
The eigenvalues of the matrix $\Delta$ are $1-\rho_1$ and
$1-\rho_2$.

The case of a single preserved SUSY is reached if only one of the
moduli of the nonzero elements in the diagonal matrix $C^\prime$
is equal to 1, for definiteness let it to be $\rho_1$,
$\rho_1=1$. After the diagonalization of the matrix $C$ the
eq.~(\ref{matk})
$$
\Delta^\prime \,{\cal K}^\prime =0
$$
requires vanishing of all entries in ${\cal K}^\prime$ except for
${\rm Im}e^{i\varphi_1 /2}\kappa^\prime_1 =\nu$ which is
arbitrary. This value plays a role of the parameter of the single
unbroken SUSY. Further,  for the
$\kappa$--symmetry parameters~(\ref{loctr}) one has
$$
\kappa = U^{-1}\left( i\nu e^{-i\varphi_1 /2} \atop 0 \right) \, .
$$
where $U$ is a unitary unimodular matrix diagonalizing the matrix
$C_{ab}$.

Thus we have shown that the model of the massive superparticle
described by the twistor--like action~(\ref{act}) possesses one or
two independent local $\kappa$--transformations which correspond
to BPS configurations preserving $1/4$ or $1/2$ of the
target--space supersymmetry. The case with $3/4$ unbroken
supersymmetry is not realized in the massive case of the
presented model.

\section{Constraints of the model}

Phase space of the model  is  parametrized by the coordinate variables
\begin{equation}
q^A = (x^\mu , y^{\alpha\beta}, \bar y^{\dot\alpha\dot\beta};
\theta^{\alpha}, \bar\theta^{\dot\alpha}; v_\alpha^a, \bar v_{\dot\alpha{}a})
\end{equation}
and by corresponding canonically conjugate momenta
\begin{equation}
p_A =
(p_{\mu}, z_{\alpha\beta}, \bar z_{\dot\alpha\dot\beta}; \pi_{\alpha},
\bar\pi_{\dot\alpha}; \omega^{\alpha}{}_a, \bar \omega^{\dot\alpha}{}^a ) \, .
\end{equation}
We take the standard definition  of the Legendre transformation
$p_A =\partial_r L/\partial q^A$ and of the graded Poisson brackets
$\left\{ q^A , p_B \right\} =\delta^A_B$ for all basic phase
variables.

The Lagrangian~(\ref{Lagr}) is homogeneous with respect to all
velocities, therefore the expressions for all momenta lead to the
primary constraints
\begin{eqnarray}
D_\alpha &\equiv& -i\pi_{\alpha}
-P_{\alpha\dot\beta}\bar\theta^{\dot\beta} - \theta^{\beta}
Z_{\beta\alpha} \approx 0 \, ,\\ \nonumber
\bar D_{\dot\alpha} &\equiv& \overline{(D_\alpha)}=-i\bar\pi_{\dot\alpha} -
\theta^{\beta}P_{\beta\dot\alpha}
-\bar Z_{\dot\alpha\dot\beta}\bar\theta^{\dot\beta}  \approx 0 \, ;\\
T_{\alpha\dot\beta}& \equiv& p_{\alpha\dot\beta} - P_{\alpha\dot\beta}
\approx 0 \, ;\\
\nonumber
R_{\alpha\beta} &\equiv& z_{\alpha\beta} - Z_{\alpha\beta} \approx 0 \, ,\\
\bar R_{\dot\alpha\dot\beta}& \equiv& \bar z_{\dot\alpha\dot\beta}
- \bar Z_{\dot\alpha\dot\beta} \approx 0 \, ;\\
\omega^{\alpha}{}_a &\approx& 0 \, ,\quad
\bar \omega^{\dot\alpha}{}^a \approx 0
\end{eqnarray}
where $P_{\alpha\dot\beta}$, $Z_{\alpha\beta}$,
$\bar Z_{\dot\alpha\dot\beta}$ have the expressions~(\ref{P})-(\ref{barZ})
in terms of the bosonic spinors. In adition, the whole system of
constraints includes the kinematic constraints
\begin{equation}
v^{\alpha{}a}v_{\alpha{}a} -2m  \approx 0 \, ,\quad
\bar v_{\dot\alpha{}a}\bar v^{\dot\alpha{}a} -2m \approx 0
\end{equation}
which are explicitly introduced into the action. The kinematic
constraints are secondary ones if Lagrange  multipliers are
assigned to canonical variables. Any other constraints do not
appear in the model.

The analysis of the $\kappa$--symmetry is based on the
consideration of the odd constraints. Their Poisson bracket
algebra is
\begin{equation}
\left\{D_\alpha\, , \bar D_{\dot\beta}\right\}=2iP_{\alpha\dot\beta} \, ,
\end{equation}
\begin{equation}
\left\{D_\alpha\, , D_\beta \right\}=2iZ_{\alpha\beta} \, ,
\end{equation}
\begin{equation}
\left\{\bar D_{\dot\alpha}\, , \bar D_{\dot\beta} \right\}=
2i\bar Z_{\dot\alpha\dot\beta} \, .
\end{equation}
The analysis of the constraints is simplified when they are
projected on the spinors $v_\alpha^a$, $\bar v_{\dot\alpha{}a}$.
For the fermionic constraints we get
\begin{equation}
D^a \equiv v^a D =
-iv^a\pi +m(\bar v^a\bar\theta -C^{ab}\theta v_b) \approx 0 \, ,
\end{equation}
\begin{equation}
\bar D_a \equiv \bar D \bar v_a= -i\bar \pi\bar v_a -
m(\theta v_a -\bar C_{ab}\bar v^b\bar\theta ) \approx 0 \, .
\end{equation}
Due to the kinematic constraints~(\ref{kinem}) the  canonically
conjugate momenta for the Grassmann variables
$\theta_a\equiv\theta v_a$ and
$\bar\theta^a\equiv -\bar v^a\bar\theta =\overline{(\theta_a)}$
are $\pi^a =v^a\pi /m$, $\bar\pi_a =\bar\pi\bar v_a /m$ and
$\left\{\theta_a ,\pi^b \right\}=
\left\{\bar\theta^b ,\bar\pi_a\right\}=\delta^b_a$.
In terms of these variables the fermionic constrains acquire simple  form
\begin{equation}
D^a =-m(i\pi^a +\bar\theta^a +C^{ab}\theta_b )\approx 0 \, ,
\end{equation}
\begin{equation}
\bar D_a =\overline{(D^a)}=-m(i\bar\pi_a +\theta_a +\bar C_{ab}\bar\theta^b )
\approx 0 \, .
\end{equation}
Their Poisson brackets are
\begin{equation}\label{brac}
\left\{D^a\, , \bar D_b\right\}=2im^2 \delta_a^b \, ,
\end{equation}
\begin{equation}
\left\{D^a\, , D^b \right\}=2im^2 C^{ab} \, ,
\end{equation}
\begin{equation}
\left\{\bar D_{a}\, , \bar D_{b} \right\}= 2im^2\bar C_{ab} \, .
\end{equation}

The algebras of the Lorentz--spinor constraints
$D_{\alpha}$, $\bar D_{\dot\alpha}$ and of the Lorentz--scalar constraints
$D^a$, $\bar D_{a}$ are identical. But in the second case
the role of the central charges is played by the Lorentz--scalar
quantities
$C^{ab}$, $\bar C_{ab}$ instead of $Z_{\alpha\beta}$,
$\bar Z_{\dot\alpha\dot\beta}$ and by the static momentum
$p^0 =m$, ${\bf p} =0$ instead of the usual four--momentum.
The consideration in terms of quantities with indices
$a$, $b$, ... is Lorentz covariant
due to the use of the bosonic variables $v_\alpha^a$ which play
the role of harmonic variables~\cite{GIKOS}-\cite{ZimF}
parametrizing an appropriate homogeneous subspace of the Lorentz
group.

The matrix of the Poisson brackets~(\ref{brac})
\begin{equation}
\left(
\begin{array}{cc}
\left\{\bar D_a\, , D^b\right\} & \left\{\bar D_{a}\, , \bar D_{b} \right\}\\
\left\{D^a\, , D^b \right\} & \left\{D^a\, , \bar D_b\right\}
\end{array}
\right) =2im^2 \Delta \,  .
\end{equation}
is in fact the matrix $\Delta$. Its eigenvalues and odd
eigenvectors have been found above. Thus,  the separation of the
first and second class Fermi constraints can be done
straightforwardly.

It is convenient to introduce the new constraints
$$
\Delta_{(\lambda)} =X_{(\lambda)} \Delta
$$
where
$D = \left( D^a \atop \bar D_a \right)$, and
$X_{(\lambda)} = \left( X^a \atop \bar X_a \right)$ is an even normalized
eigenvector of the matrix $\Delta$ with an eigenvalue $\lambda$
i.e.
$\Delta X_{(\lambda)}=\lambda X_{(\lambda)}$.
The eigenvectors with different eigenvalues are orthogonal, the
eigenvectors having the same eigenvalue can be chosen orthogonal.
Here we do not need to distinguishing the special notation of
different eigenvectors corresponding to the same eigenvalue. The
algebra of new constraints takes a very simple form
$$
\left\{\Delta_{(\lambda)}\, , \Delta_{(\lambda^\prime)}\right\}
= 2im^2\lambda \delta_{\lambda\lambda^\prime}\, .
$$

Thus the repetition of the analysis which was made in the
previous section allows us to obtain the full system of
orthonormal eigenvectors
$X_{(\lambda)}$ of the  matrix $\Delta$ and to construct the first
class constraints $D_{(0)}$ which correspond to the  zero
eigenvalues and generate the $\kappa$--symmetry transformations.

\section{Conclusion}

In this paper we have constructed the  twistor--like model of the
superparticle with tensorial central charges. The proposed model
uniformly describes cases of massive and massless superparticles.
For the description of the degrees of freedom associated with
tensorial central charges we have used coordinates of central
charges as well as additional bosonic spinors. The latter
variables have also been used for the twistor--like representation
of the momentum. In the case of zero mass one can obtain the
twistor--like formulation of the superparticle with tensorial
central charges preserving $1/2$ and $3/4$ of target--space
supersymmetry. In the massive case our model has one or two
$\kappa$--symmetries and preserves $1/4$
or $1/2$ of the target--space supersymmetry. The additional
bosonic spinors have been used as Lorentz harmonic variables.
This allowed us to eliminate auxiliary and gauge degrees of
freedom without the violation of the Lorentz invariance.

\medskip
We would like to thank I.A.Ban\-dos, E.A.Iva\-nov, S.O.Kri\-vo\-nos,
J.Lu\-kier\-ski, A.Yu.Nur\-ma\-gam\-be\-tov, D.P.So\-ro\-kin,
A.A.Zhel\-tu\-khin for interest to the work and for many useful discussions.
The authors are grateful to I.A.Ban\-dos, V.P.Be\-re\-zo\-voj,
A.Yu.Nur\-ma\-gam\-be\-tov, D.P.So\-ro\-kin
for the hospitality at the NSC Kharkiv Institute of Physics and Technology.
SF would like to thank V.V.Yanovsky for warm hospitality at the
Institute for Single Crystals of Kharkiv.

\end{document}